**Bridging the Gap between an Isolated Nanochannel/pore System and Communicating Multipore Heterogeneous Membrane System**

Yoav Green, Sinwook Park and Gilad Yossifon[*]

Faculty of Mechanical Engineering, Micro- and Nanofluidics Laboratory, Technion–Israel Institute of Technology, Technion City 32000, Israel

To bridge the gap between single/isolated pore systems to multi-pore systems, such as membranes/electrodes, we studied an array of nanochannels with varying interchannel spacing that controlled the degree of channel communication. Instead of treating them as individual channels connected in parallel or an assembly like a homogeneous membrane, this study resolves the pore-pore interaction. We found that increased channel isolation leads to current intensification whereas at high voltages electro-convective effects control the degree of communication via suppression of the diffusion layer growth.

Understanding ion transport processes through a heterogeneous permselective medium (membranes and nanoslots) is of great importance in realizing optimal designs of desalination, bimolecular sensors and fuel cell devices [1,2]. In such systems, under the application of an external electric field, the ion-permselectivity symmetry breaking phenomenon results in ionic concentration-polarization (CP), i.e. the formation of ionic concentration gradients. In this work we shall elucidate on the effect of the geometric heterogeneity and field focusing effect in a nanoslot array on the CP and its associated diffusion limited current. Thus far, nanochannel array systems that were previously investigated [3–5] treated the collective behavior as the sum of single and isolated nanochannels. In this work, we shall show that neighboring channels interact such that the collective behavior also includes communication effects that are geometry dependent. Understanding these effects in single and multiple channels/pores is of much importance to a wide variety of multi-pore structures, e.g. fabricated nanochannel array [3–5], membranes [6] and hierarchically micro-/nano-porous electrodes [7]. The fabricated nanoslot array serves as a simple and tractable model of the more geometrically complex ion-permselective membrane systems, thus, allowing us to study the role of heterogeneity.

It is well established [8,9] that the current-voltage curve (I-V) of permselective systems exhibits three regimes: Ohmic, limiting-current and over-limiting-current (OLC). In the low-voltage (i.e. under-limiting) regime, the current increases linearly with the applied voltage, in conjunction with ionic depletion at the micro-nanochannel interface through which the counterions are entering. Theoretically, under the assumption of local electroneutrality (LEN), the system cannot sustain currents larger than a limiting value [8] corresponding to the complete depletion of the salt concentration at the microchannel-permselective medium interface. In practice, measurements showed that currents could surpass this theoretically predicted value. An early work [9] showed that by waiving the LEN assumption, an extended space charge layer (SCL) is formed at the entrance of the permselective interface, resulting in large yet finite differential resistance within the limiting current regime. Other possible mechanisms responsible for the increase of the conductivity in this regime are related to the charged microchannel walls through surface conductance [10], electro-osmotic flow (EOF) [11,12], Taylor dispersion [13], and also surface charge regulation [14].

Two electro-convective (EC) mechanisms, associated with the emergence of the SCL have been shown to be responsible for the OLC are Dukhin's EOF of the second kind [15] and Rubinstein and Zaltzman's instability [16,17]. In both mechanisms it is the tangential component of the electric field that acts on the SCL to form an electric body force that drives the vortex pairs. However, the former is commonly associated with geometrically heterogeneous systems wherein a tangential field always exists. In contrast, the latter is associated with homogenous systems wherein the tangential component of the field is vanishingly small, thus, resulting in quiescent flow conditions. Rubinstein and Zaltzman [16,17] showed that beyond a critical voltage the SCL losses its stability and forms an array of EC vortices.

The formation of these vortices suppress the otherwise unbounded diffusive growth of the depletion region, which in turn, results in a shorter selected diffusion layer (DL) length that is responsible for the transition to OLC. These have been observed in a number of recent experiments for wide nanoslot [18], homogeneous membrane [19], and an heterogeneous array of narrow nanoslots [3,4]. The interplay between these two EC mechanisms has recently been studied numerically for geometrically modulated membrane surface [20] and observed experimentally for a wide nanoslot [21]. However, we here focus on the effect of geometric heterogeneity of the nanoslot interface on the ion transport in the Ohmic, limiting current and OLC regimes.

A number of recent theoretical works have studied the effects of geometric heterogeneity in permselective systems undergoing CP in 2D [22,23] and 3D [24,25], however, these effects have not been verified directly in experiment. These theoretical works focused on CP under the LEN assumption and thus neglected the effects of both the SCL and EC. Here we shall experimentally substantiate, for the first time, these theoretical predictions regarding the effect of geometric heterogeneity on CP, and in turn, the diffusion limited current. Furthermore, our experimental results extend also into the OLC regime where EC effects become significant.

The microfabricated device consists of an array of identical nanoslots of varying interchannel spacing (see FIG. 1 and Table 1) flanked between rectangular microchambers. Details on the fabrication process and the measurement technique are found in the Supplementary Material [26].

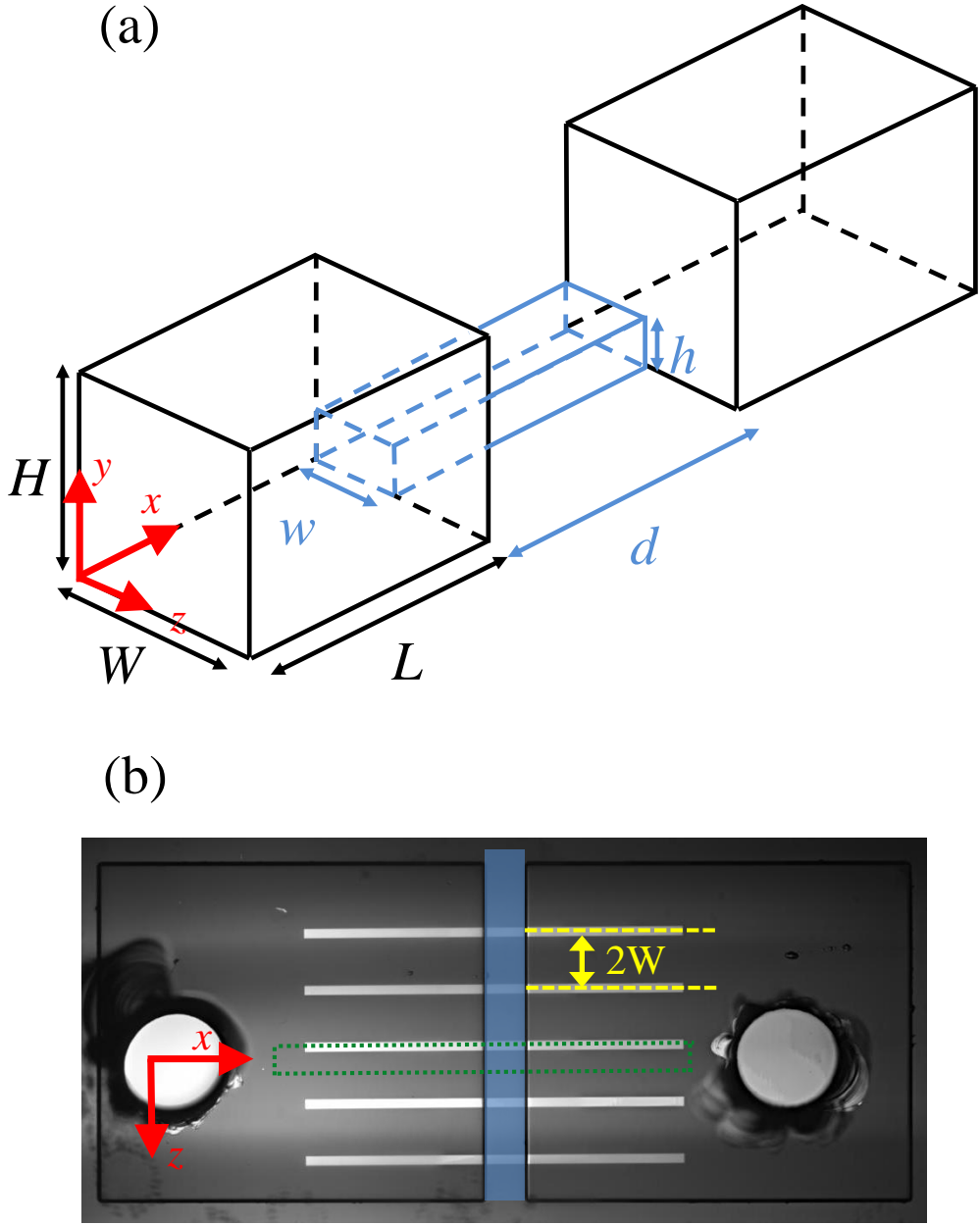

**FIG. 1.** (Color online) (a) Schematics describing half a 3D periodic unit cell consisting of a straight permselective medium connecting two opposite symmetric microchambers. (b) An optical top-view (*x-z* plane) image of a nanoslot array (half unit cell is marked by a dotted green box while the yellow dashed lines mark the distance between the center of two adjoining channels, $2W$). The microchamber has a height of $H = 48\,\mu m$ and width of $\sim 3\,mm$. The approximate distance of the drilled holes from the permselective interface is $L \sim 2\,mm$. The nanoslots have a width of $2w = 92\,\mu m$, height $h = 178\,nm$, and length $d = 0.35\,mm$. The number of channels and their spacing varies as given in Table 1. Note the depiction in (a) is not to scale.

**Table 1. Half periodic unit cell width, $W$, total number of array channels, $N$, and geometric ratios as given by Eq.(1)**

| Nanoslot array name | $W\,[\mu m]$ | $N$ | $\dfrac{L}{HW}\left[\dfrac{10^5}{m}\right]$ | $\dfrac{\bar{f}}{L}\left[\dfrac{10^4}{m}\right]$ |
|---|---|---|---|---|
| Array A | ~1500 | 1 | 0.28 | 6.00 |
| Array B | 285 | 5 | 1.46 | 4.87 |
| Array C | 200 | 7 | 2.08 | 4.65 |
| Array D | 165 | 7 | 2.52 | 4.51 |
| Array E | 93 | 7 | 4.48 | 4.15 |
| Array F | 93 | 13 | 4.48 | 4.15 |

After cleaning of the channels [26], a KCl solution of low concentration $30\mu M$ was introduced to ensure EDL overlap (Debye length of $\lambda_D \approx 55 nm$) [27]. For each configuration (Table 1) ~7-8 I-V sweeps were conducted between 0 to $50V$ at increments of $0.25V$ and a time step of $10s$. The mean value and standard deviation of these measurements are calculated for each channel separately (Fig. S1 of [26]). FIG. 2a depicts the measured mean current for each nanoslot array configuration. Expectedly, the current increases with increasing number of channels as the total Ohmic resistance of the nanoslot array (not including the microchambers and field focusing resistances [25]) decreases according to $R_{array} = R_{unit\,cell}/N$, where $R_{unit\,cell}$ is the resistance of the periodic unit cell (FIG. 1), as these are connected in parallel.

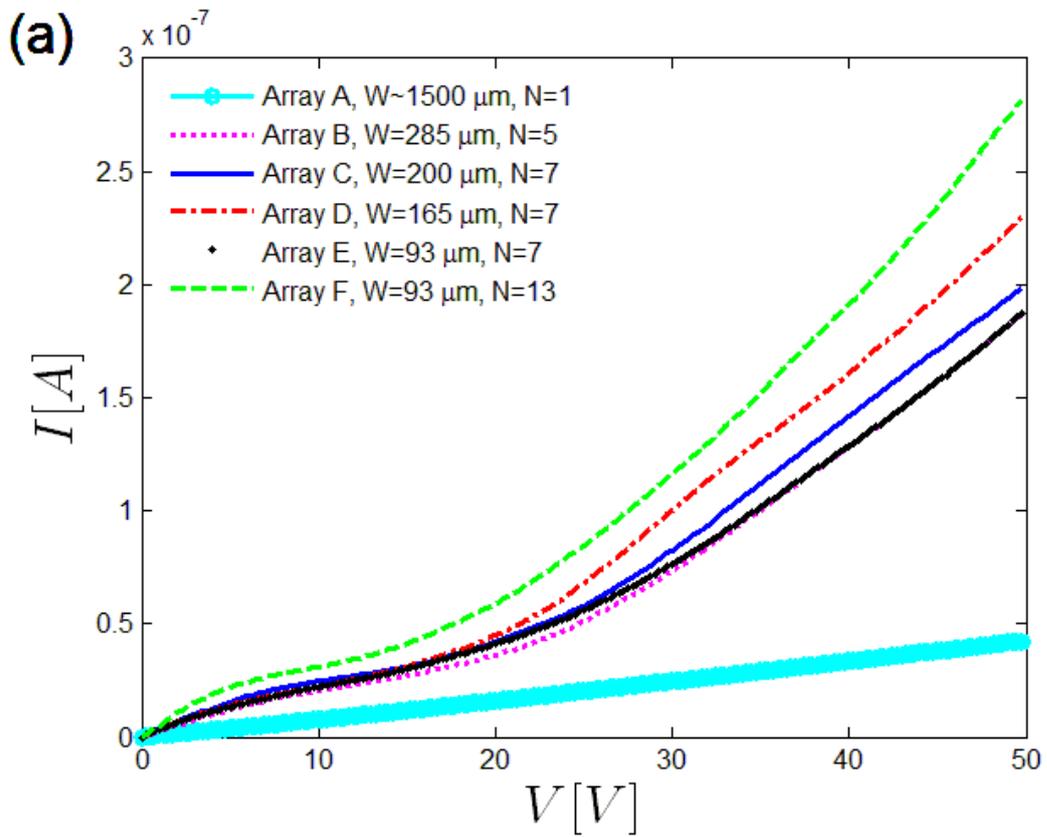

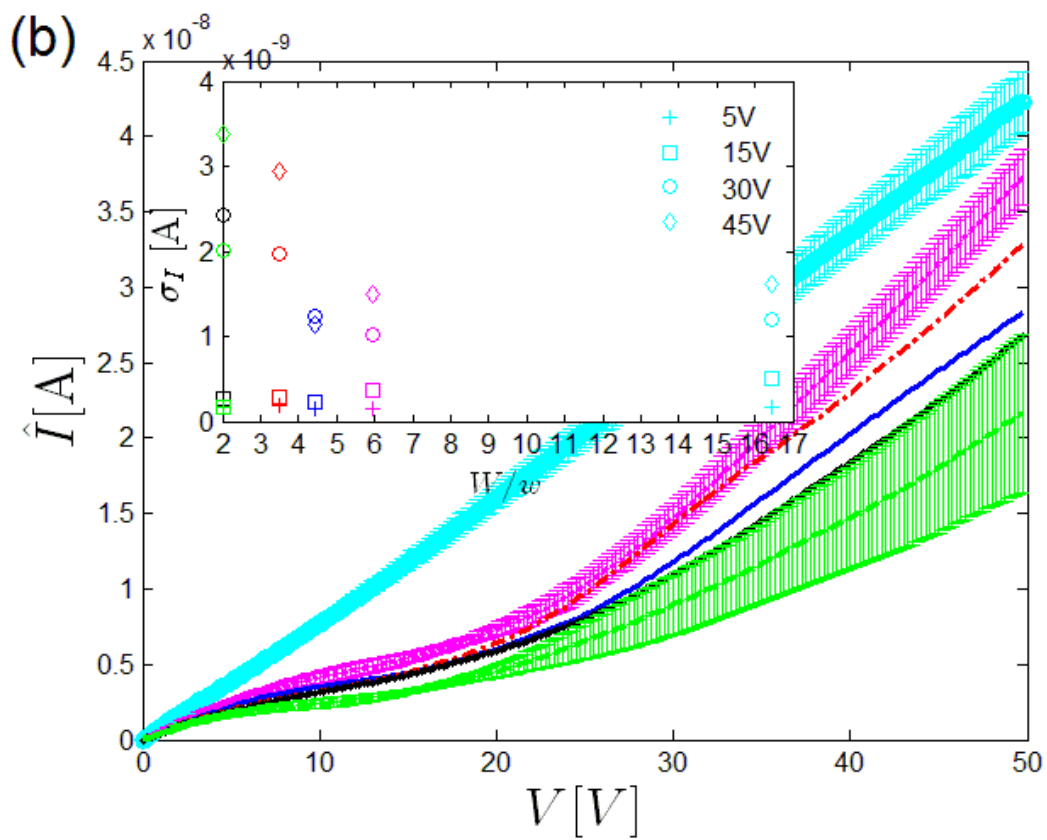

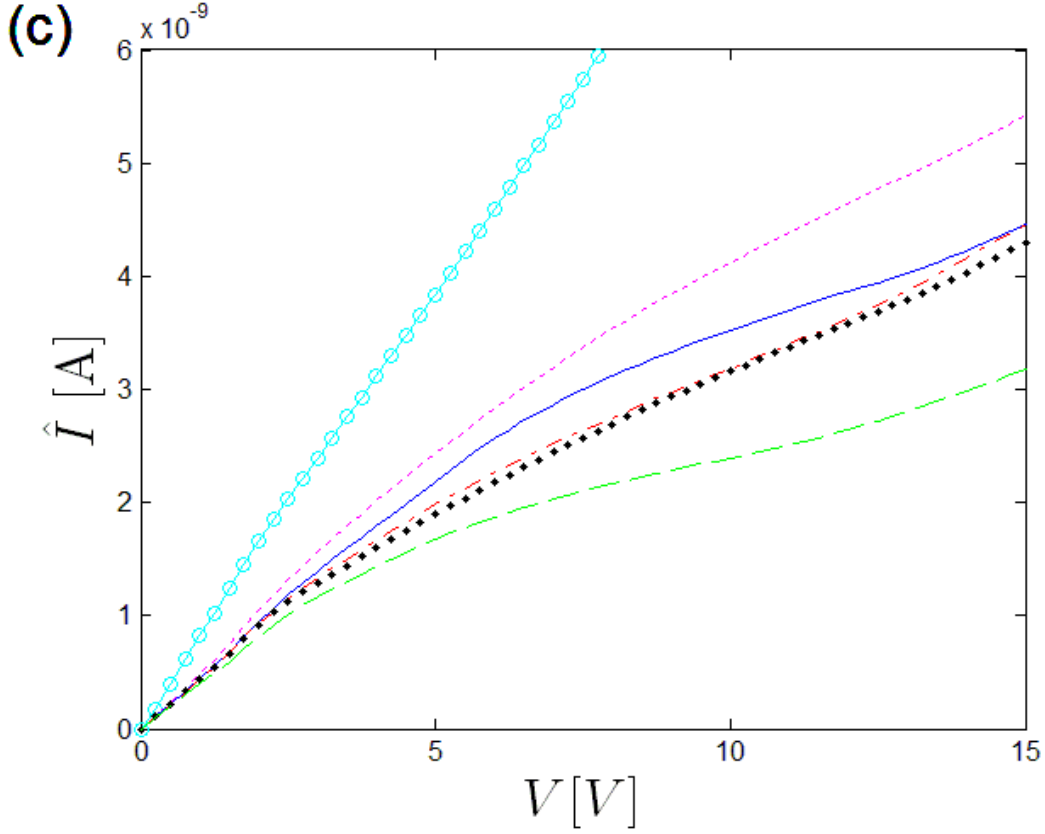

FIG. 2. (Color online) I-V curves for channels of varying interchannel spacing depicting: (a) The total average current, $I$. (b) Average current per channel, $\hat{I} \equiv I/N$. For clarity purposes a single standard deviation, $\sigma_I$, is demonstrated only for arrays A, B and F. The inset shows the calculated standard deviation at various applied voltages as a function of the periodic unit cell half-width, $W$, normalized by the nanoslot half-width, $w$. The color scheme follows that of the legend given in (a). (c) A close-up of the Ohmic and limiting current regimes of (b).

Previous works have focused on microchamber-nanochannel systems with dominating nanoslot resistance [5,27,28]. In a more recent work, a model system similar to our experimental setup, comprising of a periodic array of nanoslots, was studied theoretically [25] for the case of comparable microchannel and nanochannel resistances. In contrast to a recent paper [29] in which the ionic transport through an array of nanopores was studied under conditions of vanishing permselectivity, the former study [25] accounts for the nanoslots ionic permselectivity and the resulting CP. It was shown [25] that the total Ohmic conductance of an ideal permselective system was comprised of three components according to the system geometry (FIG. 1): nanochannel resistor (1st term), microchamber resistor (2nd term) and geometric field-focusing resistor (3rd term)

$$\sigma = \frac{\hat{I}}{V} = \frac{I/N}{V} = 2\frac{DF^2 c_0}{RT}\left(\frac{d}{hw\Sigma} + \frac{2L}{HW} + 4\frac{\overline{f}}{L}\right)^{-1}, \qquad (1)$$

where $D$ is the ionic diffusion coefficient, $R$ is the universal gas constant, $T$ is the absolute temperature and $F$ is the Faraday constant. Herein, $\Sigma = \tilde{\Sigma}/c_0$ is the excess counterion concentration (normalized by the bulk concentration $c_0$) exactly compensating the (negative) surface charge within the nanoslot and $\overline{f} = \overline{f}(L,H,h,W,w)$ is a geometric non-dimensional function with a strong dependence on the two heterogeneity parameters $h/H$ and $w/W$ (the full form of the $\overline{f}$ function can be found in Eq. 26 of Ref [25]). In the current contribution, the former ratio has been kept constant and the latter has been varied. In addition it was shown that $\overline{f}$ is a monotonically increasing function of decreasing $w/W$ and equals zero when the system is homogeneous $(w=W)$. Hence, the microchamber (2nd term) and the field focusing (3rd term) resistors in Eq.(1) counteract each other with the change of the system geometric heterogeneity (i.e. $W/w$, while keeping $w$ constant), with the former decreasing with increasing $W/w$ and vice versa for the latter (Table 1).

In the case of nanochannel dominated resistance one would expect that the current per channel versus voltage curves should collapse onto each other. FIG. 2b,c depict the current per channel (i.e. $\hat{I} = I/N$) for the voltage range of 0-50V and 0-15V, respectively. It is clearly seen that the conductance (i.e. slope of the I-V curves) per channel is not identical but rather depends on the interchannel spacing (specifically, increasing with increasing $W$). This is a consequence of the fact that the combined contribution of the microchannel and geometric field-focusing resistances in our device are comparable to that of the nanochannel.

The first term in Eq. (1) can be evaluated based on the range of values $\Sigma = \tilde{\Sigma}/c_0 = \left[(-2\sigma_s/h)/F\right]/c_0 \approx 27-233$ (wherein $\sigma_s$ is the surface charge density and was evaluated as $\sigma_s = -7 \cdot 10^{-3} \, C/m^2$ in Ref.[26] and $\sigma_s \approx -6 \cdot 10^{-2} \, C/m^2$ in Refs. [5,27]) yielding $d/hw\Sigma = 1.8 \cdot 10^5 - 1.6 \cdot 10^6$. The calculated resistance of an array of nanoslots is simply $R_{array} = R_{unit\,cell}/N = 1/\sigma N$ which for example in the case of a single nanoslot $(N=1)$ yields $R_{array}|_{N=1} \sim 1-4.2 \, G\Omega$. This is consistent with the experimentally measured value based on the slope in the Ohmic region $R_{array}|_{N=1} \sim 1.2 \, G\Omega$. Another consequence of heterogeneity is the limiting current dependence on the microchamber geometry [24,25]

$$\hat{I}_{\lim} = 2DFc_0 \left( \frac{L}{2HW} + \frac{\bar{f}}{L} \right)^{-1}, \qquad (2)$$

which is evident in FIG. 2c.

We confirm these predictions in the following manner. Three arrays with the same number of nanoslots but varying interchannel spacing (Arrays C, D and E) were measured and are compared with the single channel (Array A). As predicted from Eq.(1) the Ohmic conductance decreases with decreasing heterogeneity (i.e. decreasing $W/w$) such that $\sigma_A > \sigma_C > \sigma_D > \sigma_E$. An additional test group are Arrays B, C and F whose total *active* array width, $\sim 2WN$, is approximately the same. Array F which is comprised of the largest number of nanoslots, expectedly, exhibits the highest total current, yet, due to its relatively small heterogeneity (i.e. small $W/w$ resulting in small field focusing resistor but large microchamber resistor) it has the lowest conductance and limiting current-per channel ( FIG. 2c). Following Array A (single channel), Array B which is the most isolated (multiple) channel has the highest current per channel. The differences between Arrays B, C and F become even more pronounce when the current shifts from the Ohmic to the limiting regime when the microchamber geometry at the depleted side becomes increasingly important. Additionally the average current density per channel $\hat{I}/W$ exhibits (Fig.S2 of [26]) the expected reversal relative to the average current $\hat{I}$ [24].

The limiting currents of all channels as a function of geometric heterogeneity ($W/w$), excluding the single channel, are compared in FIG. 3. We attribute the fact that the single channel case doesn't exhibit a limiting current voltage window due to its strong isolation that results in a large tangential electric field which drives a single pair of EO vortices of the second kind (Dukhin mechanism [15]). The more efficiently mixed depleted region eliminates the saturation of the diffusion-limited current which is in agreement to a previous study of varying nanoslot width [28]. We approximate the value of the limiting current to be that corresponding to a minimal differential conductance (Fig. S1 [26]). To facilitate comparison between measured and theoretically predicted limiting currents (Eq.(2)) we normalize both to that of Array F (Table 1), yielding fair agreement between theory and experiments. We note a peculiarity with Array E – it is larger than that of Array F although both have the same interchannel spacing. This is probably due to the fact that Array E's fewer nanoslots $(N=7)$ are located at the center of the device whereas in Array F the nanoslots $(N=13)$ are less centralized, thus, their effective distance from the reservoir (see FIG. 1) is larger, which in turn, results in a smaller limiting current.

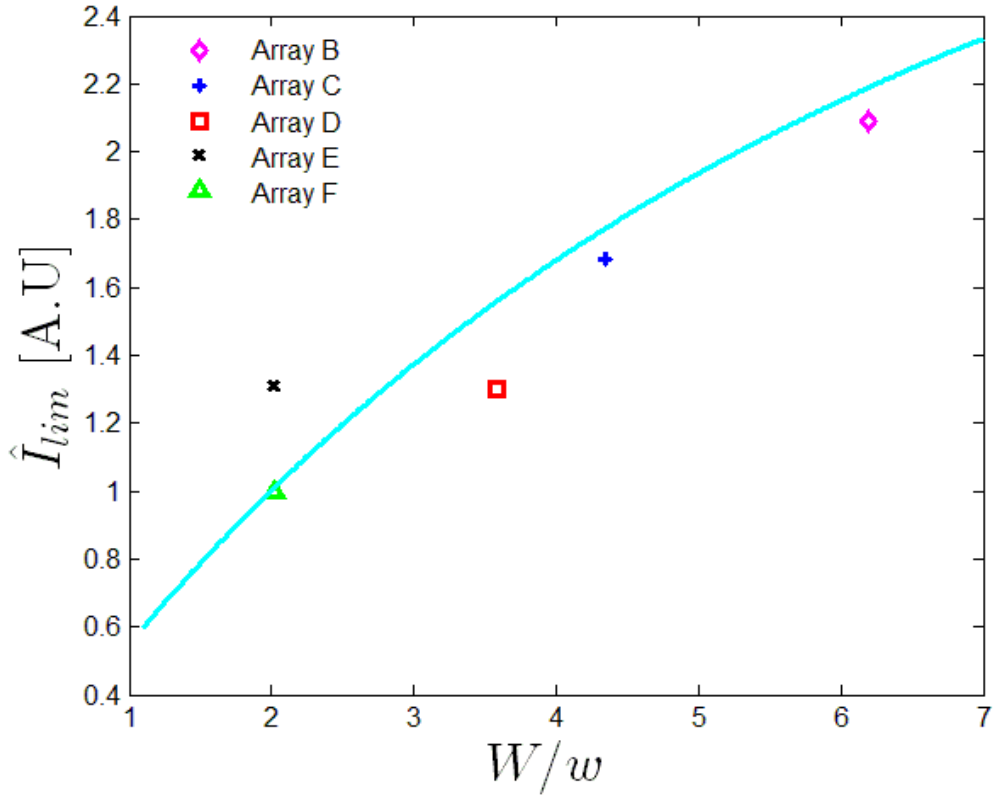

**FIG. 3 (Color online) Limiting currents (per channel) normalized by that of Array F versus system heterogeneity, $W/w$. Symbols represents experimental data, continuous line represents theoretical model (Eq.(2)).**

We finally wish to address the effects of EC. In the Ohmic and limiting current regimes the effects of EC are small, regardless of the heterogeneity ($w/W$), as is evident by the small noise (standard deviation) measured for all arrays (inset of FIG. 2b). Since the EC effects are non-linearly dependent on the applied voltage, these become the dominating mechanism for controlling the current at the OLC regime (see Supplementary Materials [26] for movies demonstrating strong convection effects). Accordingly, the noise is monotonically increasing with increased voltage (inset of FIG. 2b). FIG. 4 shows the time evolution of Array F under an applied voltage of 45V and also the images of Arrays A, B, and F after 200 seconds. In the OLC regime (inset of FIG. 2b) we can clearly see that the measured standard deviations get smaller with increasing channel isolation (i.e. high heterogeneity) and vice versa. An isolated channel has a distinct single vortex pair, which in turn, suppresses the diffusive growth of the depletion layer resulting in a smaller selected diffusion layer length (Movies 1 and 2 [26]) and significantly lower noise in the current. In contrast, at small interchannel spacing (Movies 1 and 3 [26]) strong interchannel communication leads to electro-convection effects, reminiscent of the complex vortex and depletion array wave length selection process due to the Rubinstein-Zaltzman instability [16,17]. It is likely that this complex and highly non-linear coupled electro-diffusion and convection process increases the measured noise (standard deviation). In recent theoretical studies [30–32] it was even shown for homogeneous systems that interaction of the EC instability induced vortices leads to a chaotic behavior

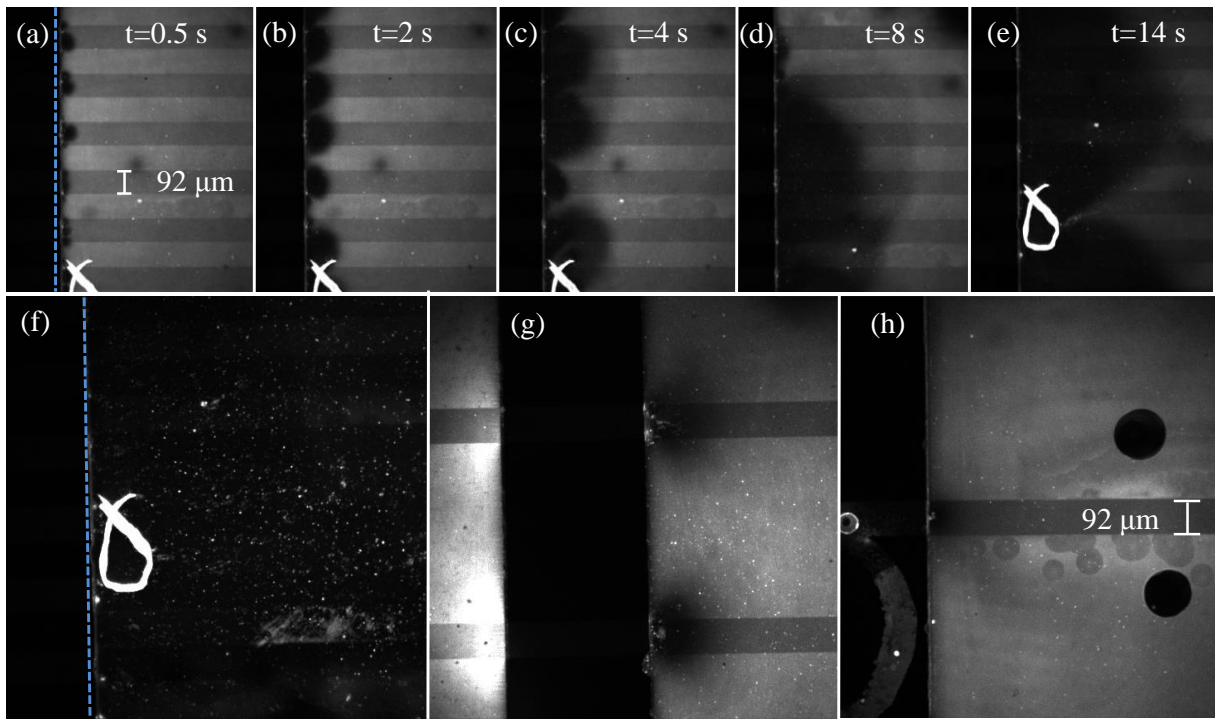

**FIG. 4 (Color online) (a)-(e) Time evolution of the depletion and vortex array of Array F under an applied voltage of 45V. (f) At steady-state, the depleted region eventually reaches the reservoirs. In contrast, for larger interchannel spacing, electroconvection slows down the diffusive growth of the depletion layer as seen in (g) for the most isolated (multiple) channels (Array B) after 200 seconds (eventually the depletion regions merge) and in (h) for the single channel (Array A) at steady state. The dashed blue line marks the interface of the nanochannel and microchambers.**

In this work we have confirmed experimentally that geometric heterogeneity (through varying interchannel spacing) of micro-nanochannel system undergoing CP not only changes the Ohmic conductance and the limiting current of the system, as was recently theoretically predicted [24,25], but also significantly changes its over-limiting response. Specifically, both the conductance and limiting current per channel increase with increasing system heterogeneity. In addition, the OLC current is also highly dependent on the geometric heterogeneity where the EC vortices can either totally suppress the diffusive growth of the depletion layer (isolated channels) or contribute to a complex interchannel communication process eventually merging into a uniform diffusive front propagating all the way to the reservoirs. Moreover, the threshold voltage for the limiting to over-limiting current transition decreases with increased heterogeneity/channel isolation (FIG. 2b). This also means that the limiting resistance voltage window vanishes with increased heterogeneity in agreement to previous studies [28,33]. However, further study with smaller interchannel spacing and channel width is required in order to test the validity of earlier predictions [26] that for a given active permselective surface there might be an optimal interchannel spacing that results in a local maxima of the OLC. Our findings, thus, serve as an initial step in understanding the full effects of geometrical heterogeneity on CP coupled with electroconvection on a breadth of communicating multi-pore structures, e.g. nanochannel array, membranes and hierarchically micro-/nano-porous electrodes. It may also be useful for optimal design of pore spacing in synthesized membranes/fabricated nanochannel array.

This work was supported by ISF Grant No. 1078/10. We thank the Technion RBNI (Russell Berrie Nanotechnology Institute) and the Technion GWRI (Grand Water Research Institute) for their financial support. YG would like to thank the Rieger Foundation for their support.